\documentstyle[12pt,procsla]{article}
\begin{document}
\centerline{\normalsize\bf SCHR\"ODINGER OPERATORS}
\baselineskip=22pt
\centerline{\normalsize\bf GENERATED BY SUBSTITUTIONS
\footnote {\footnotesize Work partially supported by EU contract
CHRX-CT93-0411} }

\centerline{\footnotesize Anton BOVIER}
\baselineskip=13pt
\centerline{\footnotesize\it Weierstrass-Institut f\"ur Angewandte Analysis und
Stochastik}
\baselineskip=12pt
\centerline{\footnotesize\it Mohrenstrasse 39, D-10117 Berlin, Germany}
\centerline{\footnotesize\it E-mail: bovier@iaas-berlin.d400.de}
\vspace*{0.3cm}
\centerline{\footnotesize and}
\vspace*{0.3cm}
\centerline{\underbar{\footnotesize Jean-Michel GHEZ}}
\baselineskip=13pt
\centerline{\footnotesize\it Centre de Physique Th\'eorique}
\baselineskip=12pt
\centerline{\footnotesize\it CNRS, Luminy Case 907, F-13288 Marseille Cedex 9}
\baselineskip=12pt
\centerline{\footnotesize\it and}
\centerline{\footnotesize\it Phymat, D\'epartement de Math\'ematiques,
Universit\'e de Toulon et du Var}
\baselineskip=12pt
\centerline{\footnotesize\it B.P. 132 - F-83957 La Garde Cedex, France}
\centerline{\footnotesize\it E-mail: ghez@cpt.univ-mrs.fr}
\vspace*{0.9cm}
\abstracts{Schr\"odinger operators with potentials generated by primitive
substitutions are simple models for one dimensional quasi-crystals. We
review recent results on their spectral properties. These include in
particular an algorithmically verifiable sufficient condition for their
spectrum to be singular continuous and supported on a Cantor set of zero
Lebesgue measure. Applications to specific examples are discussed.}

\vspace*{0.6cm}
\normalsize\baselineskip=14pt
\setcounter{footnote}{0}
\renewcommand{\thefootnote}{\alph{footnote}}

\section{\bf Introduction}
\vspace*{0.5cm}
We consider one dimensional Schr\"odinger operators $H$ defined by
\begin{equation}
H\psi_n=\psi_{n+1}+\psi_{n-1}+V_n\psi_n,\quad\psi\in l^2(\bf Z)
\end{equation}
where $(V_n)_{n\in\bf Z}$ is an aperiodic sequence generated by a substitution.

A {\it substitution} is a map $\xi$ from a finite alphabet $\cal A$ to the set
${\cal A}^*$ of words on $\cal A$, which can be naturally extended to a map
from ${\cal A}^*$
to ${\cal A}^*$ and then to a map from ${\cal A}^{\bf N}$ to ${\cal A}^{\bf
N}$. We also define the
free group $\widehat{\cal A}^*$, extension of ${\cal A}^*$ obtained by addition
of
the formal inverses of the letters in $\cal A$ as generators. $\xi$ is said
{\it primitive} if $\exists k\in\bf N$ s.t. $\forall (\alpha,\beta)\in {\cal
A}^2$, $\xi^k(\alpha)$
contains $\beta$. A {\it substitution sequence} or {\it automatic sequence} is
a
$\xi$-right fixpoint $u_r=\alpha_r...$ given by indefinite iteration of $\xi$
{}~\cite{[1]}.
By choosing a $\xi$-left fixpoint $u_l=...\alpha_l$ such that the word
$\alpha_l\alpha_r$ is contained in $u_r$, we define by concatenation a doubly
infinite word $w=u_lu_r$.
We say that a Schr\"odinger operator of type (1) is {\it generated}
by $\xi$ if the sequence $(V_n)_{n\in\bf Z}$ is defined by $V_n=v(w_n)$, where
$v$ is a map ${\cal A}\rightarrow\bf R$.

Such an operator, with the Fibonacci sequence, has become popular as a simple
model for electron transport in
one-dimensional quasi-crystals ~\cite{[2]} (cut-and-project method). The study
of this particular example ~\cite{{[3]},{[4]},{[5]},{[6]}} and others, namely
the Thue-Morse ~\cite{{[7]},{[8]}} and period-doubling ~\cite{[8]} sequences,
has led to the following common expectations for this type of operators:
\hfill\break (i) Their spectrum is purely singular continuous and supported on
a Cantor set of zero Lebesgue measure ~\cite{{[3]},{[4]},{[5]},{[7]},{[8]}}
\hfill\break (ii) The spectral gaps are labelled by a countable set of
algebraic numbers, depending on the substitution ~\cite{{[6]},{[7]},{[8]}}.

By means of the K-theory of $C^*$-algebras, (ii) has been proven to be correct
for all operators generated by primitive substitutions ~\cite{[9]}, with the
limitation that in general one cannot exclude that some gaps are closed. A
general
perturbative approach to compute the opening of gaps has been proposed in
{}~\cite{[10]}.

\vspace*{1cm}
\section{\bf Cantor spectrum}
\vspace*{0.5cm}

The basic tool to establish (i) ~\cite {[11]} is the transfer-matrix formalism.
That is, the
study of the properties of solutions of Eq. (1) leads to the analysis of
products
of two-by-two matrices of the form $P_n(E)=\prod_{k=1}^{n} T_E(w_{n-k})$,
where $w_k$ is the $k$-th letter in the substitution sequence $w$ and
$T_E:{\cal A}\rightarrow SL(2)$ is a map that assigns, for fixed energy $E$, to
each
letter in the alphabet a unimodular two-by-two matrix.  In the case of Eq. (1),
$
T_E(w_n)=\left(\matrix{E-v(w_n)&-1\cr1&0}\right)
$, but the precise form of this map is not important, and therefore the same
results hold for all types of second order equations that can be reduced to a
problem of products of unimodular $2\times 2$-matrices. This includes in
particular the continuous Laplacian with piecewise constant potential of a
finite
number of different shapes (Kronig-Penney model, see ~\cite{[12a],[12]}). The
main idea of the
proof of the singularity of the spectrum $\sigma (H)$ of $H$ is its
identification with
the set $\cal O$ of zero Lyapunov exponents
$\gamma(E)=\lim_{n\uparrow\infty}\frac{1}{n}\log||T_E^{(n)}||$, known to be of
zero Lebesgue measure by a general result in ~\cite{[8]} based on a theorem of
Kotani ~\cite{[13]}.
To do this requires a rather careful analysis of the asymptotic properties of
$P_n(E)$, which is made possible by the self-similar structure of the
potential.
Let, for any word $\omega\in{\cal A}^*$,
$T_E(\omega)\equiv \prod_{\alpha\in\omega}T_E(\alpha)$ and
\begin{equation}
T^{(k)}_E(\omega)\equiv T_E^{(k-1)}(\xi(\omega))\indent\hbox{with}\indent
T_E^{(0)}\equiv T_E
\end{equation}
{}From this recursion one can obtain an even more useful system of recursive
equations
for the traces of these transfer matrices. In general there exists a finite
subset
of words ${\cal B}\subset {\cal A}^*$ containing $\cal A$ for which Eq. (2)
yields a closed set of recursive
polynomial equations for the quantities $x_E^{(k)}(\beta)\equiv tr T^{(k)}_E(
\beta)$, $\beta\in \cal B$, which is called the {\it trace map}
{}~\cite{{[14]},{[15]},{[11]}}. It turns out that to
each trace map one can associate, keeping only the term of highest degree in
each polynom, a {\it reduced trace map} which is {\it monomial} ~\cite{[11]},
that one can consider as a substitution $\phi$ on the set of traces of transfer
matrices of elements of $\cal B$, identified with $\cal B$ in the following,
whose properties are ultimately crucial
for the spectral analysis. We call such a substitution {\it semi-primitive} if:
\hfill\break (i) There exists ${\cal C}\subset\cal B$ such that $\phi|_{\cal
C}$ is a primitive
substitution from $\cal C$ to ${\cal C}^*$;
\hfill\break (ii) There exists $k$ such that for all $\beta\in\cal B$, $\phi^k(
\beta)$ contains
at least one letter from $\cal C$.

With this notation, the main result proven in ~\cite{[11]} is the following.

\vspace*{0.5cm}
{\bf THEOREM} ~\cite{[11]}: {\it Let $H$ a one-dimensional Schr\"odinger
operator of type (1)
generated by a primitive substitution $\xi$ on a finite alphabet $\cal A$.
Assume
that its reduced trace map is associated to a semi-primitive substitution
$\phi$. Assume also that there is a $k$ such that $\xi^k(0)$ contains
$\beta\beta$
for some $\beta\in\cal B$. Then the spectrum of $H$ is singular and supported
on a set
of zero Lebesgue measure. If moreover there exist $n_0,m<\infty$ such that
$\xi^{n_0}(0)=\xi^m(\gamma_0)\Gamma\omega$, where $\gamma_0\in\cal C$,
$\Gamma,\omega\in\widehat{\cal A}^*$,
and $\Gamma=\xi^m(\gamma_0)\delta$ for some $\delta\in\widehat{\cal A}^*$, then
$H$ has no eigenvalues.
Therefore, the spectrum of $H$ is purely singular continuous and supported
on a Cantor set of zero Lebesgue measure.}
\vspace*{0.5cm}

This is in fact an algorithmic procedure to prove the singularity or the
singular
continuity of the spectrum of $H$. Note that the supplementary hypothesis for
the
second result is probably not necessary, since there is at least one example
(Thue-Morse) for which it is not satisfied although $\sigma (H)$ is singular
continuous.

\vspace*{1cm}
\section{\bf Examples}
\vspace*{0.5cm}

The hypothesis of the theorem have been shown to be satisfied in a large
number of particular cases in ~\cite{[11]}:
\hfill\break {(i)} The Fibonacci sequence: $\xi(a)=ab, \xi(b)=a$: $\sigma (H)$
is
singular continuous, as already proven in ~\cite{{[4]},{[5]}};
\hfill\break {(ii)} The Thue-Morse sequence: $\xi(a)=ab, \xi(b)=ba$: $\sigma
(H)$ is
singular as already proven in ~\cite{[8]}, where we proved that it is in
fact singular continuous;
\hfill\break {(iii)} The period-doubling sequence: $\xi(a)=ab, \xi(b)=aa$:
$\sigma (H)$ is singular continuous, as already proven in ~\cite{[8]};
\hfill\break {(iv)} The circle sequence: $\xi(a)=cac, \xi(b)=accac,
\xi(c)=abcac$:
$\sigma (H)$ is singular continuous. More recently, in ~\cite{[16]} a new class
of examples was provided by sequences derived from circle maps with rotation
numbers obtained from precious means;
\hfill\break {(v)} The binary non-Pisot sequence: $\xi(a)=ab, \xi(b)=aaa$:
$\sigma (H)$ is singular. We can only conjecture that it is singular
continuous;
\hfill\break {(vi)} The ternary non-Pisot sequence: $\xi(a)=c, \xi(b)=a,
\xi(c)=bab$:
$\sigma (H)$ is singular continuous;
\hfill\break {(vii)} The Rudin-Shapiro sequence: $\xi(a)=ac, \xi(b)=dc,
\xi(c)=ab,
\xi(d)=db$: in this case, $\phi$ is not semi-primitive and thus, presently,
we cannot even prove the singularity of the spectrum, which is not so
surprising if one recalls that it is the `most random' of all these sequences
{}~\cite{{[9]},{[10]}}.

Let us emphasize that our theorem provides explicit examples of operators with
singular continuous spectrum. Complementary results have recently been obtained
by
Hof et al.~\cite{[17]} in which for certain types of so-called ``palindromic''
substitutions it
was shown that there exists an infinite number of unspecified translates of the
original sequence for which the corresponding operator has singular continuous
spectrum.

\vspace*{1cm}
\section{\bf Concluding remarks}
\vspace*{0.5cm}

We end this note with some remarks on the nature of the solutions of the
Schr\"odinger
equation (1). The proof of the theorem shows in fact that for energies in the
spectrum,
no solution tends to zero at infinity, but these leaves room for a variety of
behaviours. However, in most cases, there are countable subsets of energies in
the spectrum,
characterized by the fact that the transfer matrices $T_E^{(n)}(\alpha)$, for
given $n$,
commute for all $\alpha\in \cal A$, at which the solutions are extended in a
very regular way
in that they consist of pieces of repeating patterns arranged according to the
substitution ~\cite{[12]}. Such solutions have actually been discovered already
in ~\cite{[7]} and ~\cite{[8]},
but they have been recently re-discovered in numerical investigations several
times and
have given rise to erroneous claims of coexisting absolutely continuous
spectrum. While
this is nonsense, it is not unlikely that these states will be quite important
for the
transport properties of systems with singular continuous spectrum, a problem
that still
has not been satisfactorily investigated.

\end{document}